\documentclass[a4paper, 11pt]{article}
\usepackage{url}

\usepackage{amsfonts} 
\usepackage{graphicx} 
\usepackage[utf8]{inputenc}
\usepackage[T1]{fontenc}
\usepackage[english]{babel}
\usepackage{indentfirst}
\usepackage{amsmath}
\usepackage{times}
\usepackage{sectsty}
\usepackage{listings}
\usepackage{subfiles}
\usepackage{setspace}
\usepackage{lipsum}
\usepackage{courier}
\usepackage[english]{babel}
\usepackage{caption}
\usepackage{subcaption}
\usepackage{xcolor}
\usepackage{tocloft}
\usepackage{capt-of}
\usepackage{float}
\usepackage[toc,page]{appendix}
\usepackage{afterpage}
\usepackage{authblk}
\usepackage{rotating}

\cftsetindents{subsection}{0.3in}{0.3in}
\cftsetindents{subsubsection}{0.6in}{0.4in}
\cftsetindents{paragraph}{1.0in}{0.6in}

\DeclareGraphicsExtensions{.pdf,.png,.jpg,}
\allsectionsfont{\textsf}

\author[1]{Emilia M. Wysocka}
\author[2]{Valery Dzutsati}
\author[3]{Tirthankar Bandyopadhyay}
\author[4]{Laura Condon}
\author[5]{Sahil Garg}

\affil[1]{University of Edinburgh, UK}
\affil[2]{Arizona State University, USA}
\affil[3]{CSIRO, Australia}
\affil[4]{Colorado School of Mines, USA}
\affil[5]{University of Southern California, USA}

\newcommand{\horrule}[1]{\rule{\linewidth}{#1}} 	
\title{
		\usefont{OT1}{bch}{b}{n}
		\horrule{0.5pt} \\[0.4cm]
		\huge Building models for biopathway dynamics using intrinsic dimensionality analysis\\
		\horrule{2pt} \\[0.5cm]
}
\date{September 25, 2015}

\begin{document}
\maketitle
\thispagestyle{empty}
\begin{figure}
\begin{center}
\includegraphics[width=0.3\textwidth]{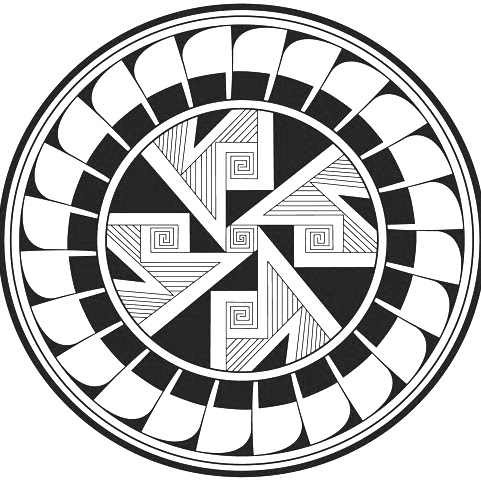}\\
\normalfont \normalsize \textsc{Complex Systems\\ Summer School 2015 \\ Santa Fe Institute\\ NM USA}
\end{center}
\end{figure}

\newpage
\tableofcontents

\newpage
\section{Introduction and project motivations}

Extensive development of technologies and methods related to data acquisition, sharing and storage have made analysis and knowledge discovery unprecedentedly challenging. For instance, big data has become increasingly common in social sciences and requires new techniques of analysis, including non linear time series approaches. One of such recent examples of a challenging dataset is the data on rebel violence in the volatile Russian North Caucasus region \cite{toft2015islamists}. The dataset has recordings of incidents of rebel violence on weekly basis for every town and village of the region. Overall, this resulted in over 1 million observations with nearly 200 variables, approximately 200 million data points. 



An important task for many if not all the scientific domains is efficient knowledge integration, testing and codification. It is often solved with model construction in a controllable computational environment. In spite of that, the throughput of \textit{in-silico} simulation-based observations become similarly intractable for thorough analysis. This is especially the case in molecular biology, which served as a subject for this study. 

In this project, we aimed to test some approaches developed to deal with the curse of dimensionality. Among these we found dimension reduction techniques especially appealing. They can be used to identify irrelevant variability and help to understand critical processes underlying high-dimensional datasets. Additionally, we subjected our data-sets to nonlinear time series analysis, as those are well established methods for results comparison. 

To investigate the usefulness of dimension reduction methods, we decided to base our study on a concrete sample set. The example was taken from the domain of systems biology concerning dynamic evolution of subcellular signalling. Particularly, the dataset relates to the yeast pheromone pathway and is studied \textit{in-silico} with a stochastic model. The model reconstructs signal propagation stimulated by a mating pheromone.

In the following sections we will elaborate on the reason of multidimensional analysis problem in the context of molecular signalling. Next, we will introduce the model of choice, simulation details and obtained time series dynamics. A description of used methods followed by a discussion of results and their biological interpretation will finalise this report. This study is a preliminary analysis of the dataset, future work will expand on the results presented here.

\section{Working with example}

\subsection{Problem view: combinatorial explosion of cell signalling systems}

As with all signal processing systems, cell signalling is characterised by signal related functionalities, such as input fidelity, output specificity, signal amplification, the sensitivity and diversity of response or the flexibility of reaction~\cite{Bardwell2007}. These highly sophisticated functions produce complex systems embodied by the combinatorial explosion of molecular interactions and states~\cite{Hlavacek2006,Chylek2014}.

On the lower level, cell signalling depends on formation and interactions of multi-subunit complexes, mainly formed by interacting proteins. They are composed from often numerous and autonomously folding blocks called domains, acting as protein functional interfaces. Importantly, protein activity is determined by multiple post-translational modification sites (phosphorylation, acetylation, ubiquitination), transitionally changing their states. For example, lets consider an ubiquitously present Epidermal Growth Factor Receptor (EGFR), which has 9 sites resulting in 512 possible states ($ 2^{9}=512$ , on- and off-state). Furthermore, each site has at least one binding partner rising the value of single receptor protein states to 19,683 possibilities ($ 3^{9}=19,683$).
The large number of possible states, even within this relatively simple system is one of the key challenges for mechanistic modelling of signalling networks. Traditional equation-based models are capable of representing only extensively studied and limited size signalling circuits. Any larger integrative models become intractable, impossible to reuse or even proofread~\cite{Lopez2013}. These problems have been addressed by rule-based modelling methods embodied by flexible languages such as Kappa~\cite{Danos2007} and BioNetGen~\cite{Faeder2009}, facilitating the creation of large and complex dynamical models.
In contrast to the other modelling techniques, in rule-based models the system emerges with time, often showing unpredictable behaviour arising from elementary reaction rules. However, their construction and analysis often limit their potential application. For instance, even though provided with visualization tools for static and causal analysis, a modeller has to resort to a self-assembled battery of tests trying to unfold the complexity of results~\cite{Suderman2013}.

\subsection{Yeast pheromone response pathway model}

In the domain of molecular biology the yeast pheromone cell cycle is an extensively studied example, both \textit{in-vivo} and as a computational model. It's often used to test hypotheses and investigate details related to mechanisms of signalling processes, such as dynamical pathway adaptation to demanding environmental conditions~\cite{Majumdar2014}, evolutionary preserved functional units (G-protein coupled receptor signalling~\cite{Dixit2014}, mitogene-activated protein kinase~\cite{Levchenko2000}), signal-noise decoupling~\cite{Dixit2014} and information transmission~\cite{Yu2008}.

\textit{Saccharomyces cerevisiae} yeast, is a model species, capable of sexually reproducing in pairs of opposite sexes (type a and $\alpha$). The mating signal is communicated by either of the cell type through pheromone release (\textit{a-factor})~\cite{Majumdar2014}. The model used in this study relates to a subcellular signalling activated in the other cell in response to the stimulus~\cite{Suderman2013}.

The pathway represents canonical mechanisms of the subcelluar signal propagation, such as G-protein activation via a GPCR, which is stimulated by pheromone ligands. The scaffold protein (Ste5) is recruited to the cell surface. Its major role is to insulate the kinase phosphorylation cascade from activating other related pathways. Ste5 dimerizes and aggregates five more proteins that phosphorylates each other forming an activation cascade. The last one is doubly activated mitogene-activated protein kinase (MAPK, Fus3) that travels to the nucleus and releases the transcription factor (TF) from its inhibitors. In this way TF transcribes genes regulating yeast mating behaviour.

The study  is examining the established hypothesis that signals in cells are propagated via well defined complexes of molecular machines rather than loosely assembled and polymorphic ensembles.

As it was shown, even though a conserved structure of decameric complex was hardly present in the ensemble model over repeated simulations, the signal was uninterrupted leading to St4 synthesis. Furthermore, contrary to the machine model, the ensemble model was able to replicate the experimental observation of combinatorial inhibition of phosphorylated Fuss3 (Fus3pp), when a copy-number of St5 was increased 60 fold. Models were built with the rule-based formalism that allows us to sample the sets of possible protein complexes the model can produce, without explicitly imposing the set of species that are formed~\cite{Suderman2013}. More details about the formalism are in the next section. The code with the models' implementation is in the public domain and can be found as one of the attached files to this paper.

		\begin{figure}
		\includegraphics[width=\textwidth]{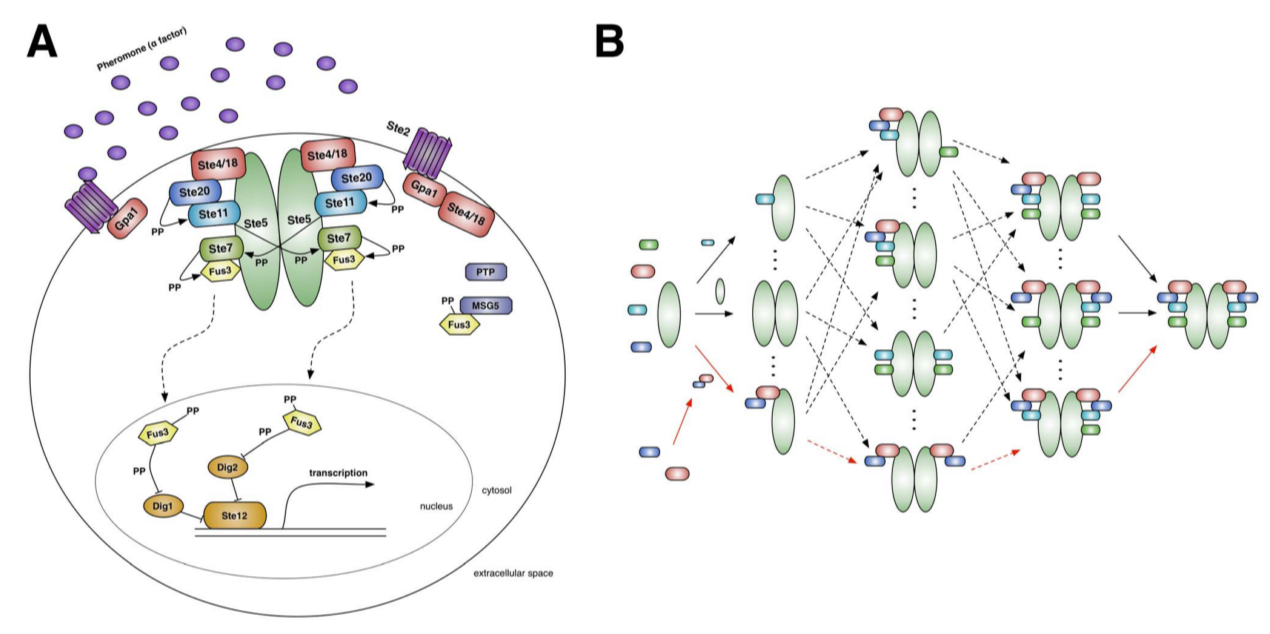}
		\caption{A: Usual scheme of hierarchically structured molecular machines B: Potential combinations of complexes appearing over the simulation. The red-arrow path represents the possible way of construction of decamer complex. Source:~\cite{Suderman2013}}
		\label{}
		\end{figure}

\subsection{Rule-based modelling}

The subject of signalling pathways and networks has already been addressed by many modelling formalisms. However, one significant advantage of the rule-based (RB) modelling is that it is able to express an infinite number of reactions with a small and finite number of rules, i.e. a single reaction rule and its parameters generalize a class of multiple interactions. In all of the other modelling methods every chemical species has to be specified in advance which is highly problematic for species with dozens of phosphorylation sites and many possible states. This limiting factor makes these methods inappropriate for modelling large-scale complex dynamical systems.

RB modelling is a method for the formal representation of combinatorially complex signalling systems in both a qualitative and quantitative way. The major idea is to replace equations with interaction rules. A rule representation is a graph-rewriting, where a graph specified on the left-hand-side is a pattern to be matched to instances in the current ``mixture'' of graphs and transformed into graphs specified on the right-hand-side. Matching should satisfy \textit{embedding}, i.e. injections on agents (graphs) with the preservation of names, sites, internal states and edges~\cite{Danos2007}.
In the rule-based language nomenclature, ``agents'' are most elementary molecules and ``species'' are agent complexes having particular states.
A model can be translated into a system of ODE equations or simulated with a stochastic algorithm. In the latter case, system trajectories are created by rule selection at each time step, which is applied probabilistically, based on reaction rates and the initial/current copy number of agents~\cite{Hlavacek2006}. Immediate consequences of this formulation are different levels of rule contextualisation (“don't care don't write”), without obligation of \textit{ad hoc} assumptions about the system, modularity, reusability and extensibility of the modelling process~\cite{Lopez2013}.
Furthermore, the ability to capture a protein as a graph with (binding) sites (e.g. domains) that have internal state(s) (e.g. phosphorylated) gives a sufficiently expressive system to capture all of the principal mechanisms of signalling processes (e.g. dissociation, synthesis, degradation, binding, complex formation~\cite{Liu2012}) as well as insight into site-specific details of molecular interactions such as affinities, dynamics of post-translational modifications, domain availability, competitive binding, causality and the intrinsic structure of interactions.

\begin{figure}
		\includegraphics[width=1.2\textwidth]{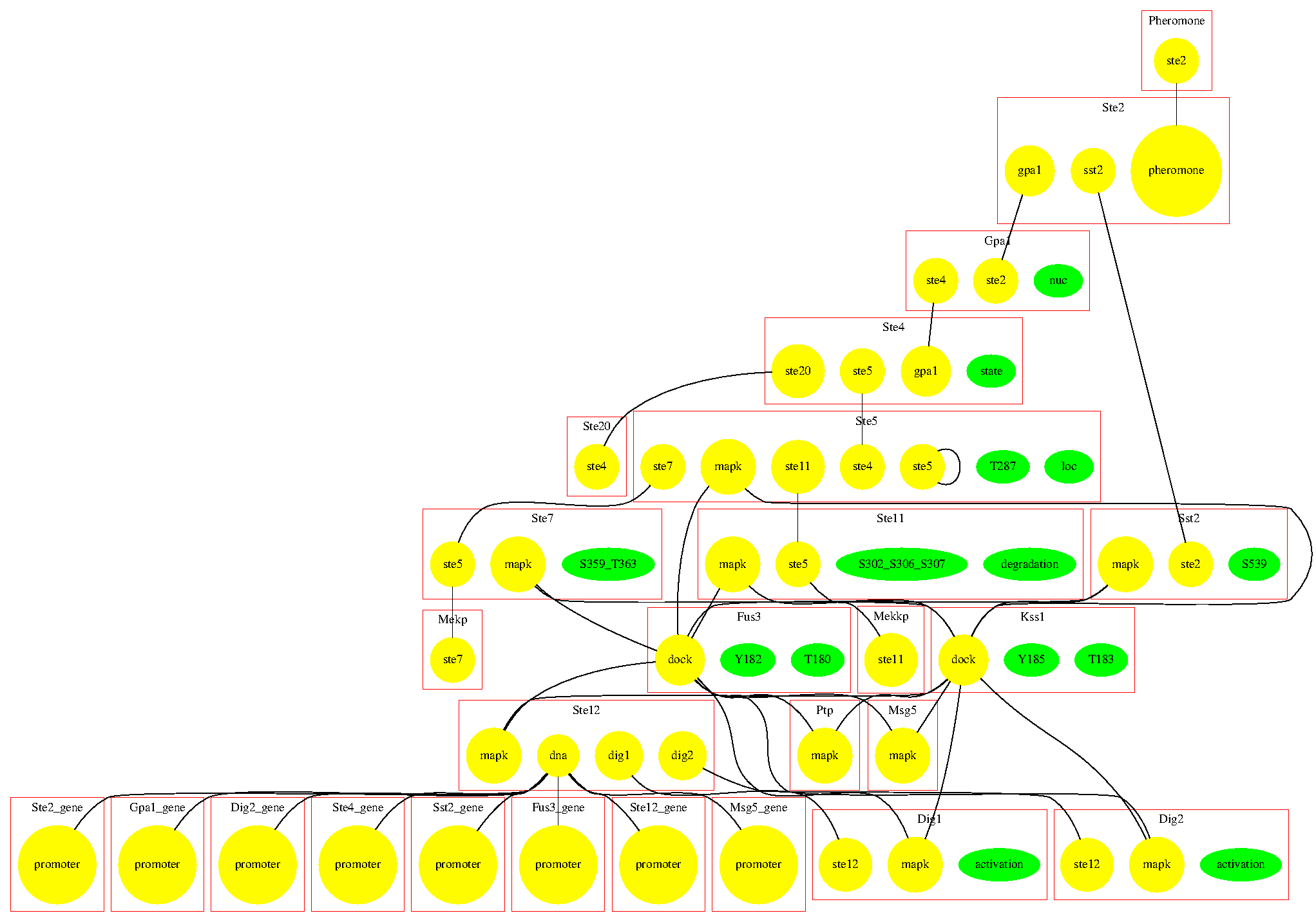}
		\caption{Contact map defined without running the simulation with KaSa software accompanying KaSim4.0 simulation tool. Yellow circles denote agents sites, green circles agent states, and edges all potential connections between species.}
		\label{fig:contactmap}
		\end{figure}

RB modelling originates from concurrency system representations and as such has the ability to capture dependencies, causality and conflicts in biological interactions (overlooked by concentration-based ODEs). In other words, precedences occurring along trajectories (stories) reveal competing events leading to a final state~\cite{Danos2007}. In the Kappa language, this feature is supported by the syntax for graphical analysis provided in the simulation tool. Among these are diagrams with causal flows, flux and influence maps as well as contact maps [Figure~\ref{fig:contactmap}] that facilitate the process of modelling. The causal flow diagram shows dependencies and conflicts in tracking indicated species and the flux maps [Figure~\ref{fig:fluxmap}], negative/positive activity transfers between rules with the quantitative contributions on edge weights, both generated on the fly during a simulation~\cite{Feret2012}. At any time of a simulation, a snapshot can be taken to record the collection of species existing at that time.

		\begin{figure}
		\includegraphics[width=1.2\textwidth]{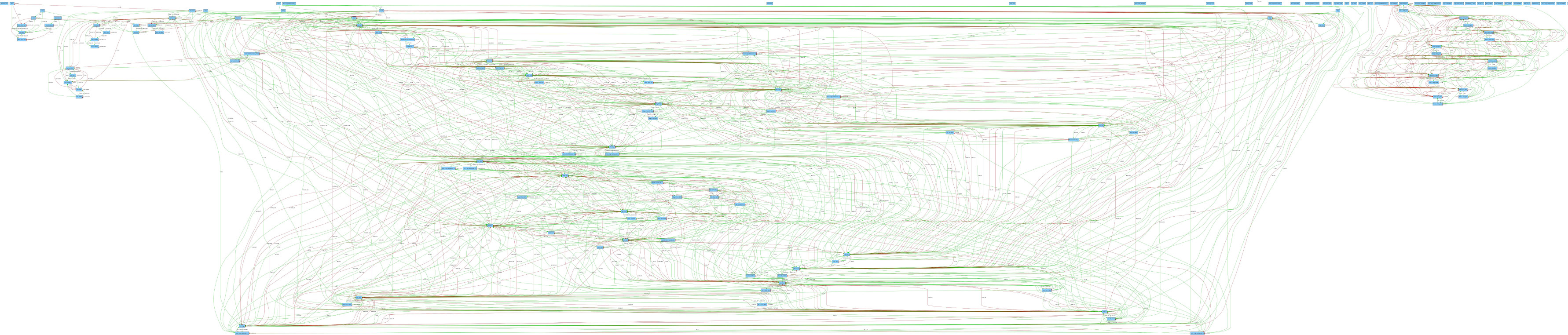}
		\caption{Flux map for pheromone pathway model in steady state simulated with KaSim4.0 }
		\label{fig:fluxmap}
		\end{figure}

\subsection{Datasets and simulations}
\label{sec:datasets}

Our time series datasets report changes of indicated molecular species' copy-number over 13,800 time points. Stochastic simulations were run for 4,600 sec with 3 time points recorded per second. The system was first simulated over 1,000 sec to reach a steady state and the initial mixture of protein complexes. Afterwords, a pheromone stimulus was introduced and the system was simulated for another 3,600 sec.

Variables in the rule-based syntax are called ``observables'' (\texttt{ \%obs:}) and are specified in a separate code block.
A single observable can be mapped to one or more rules conditioned on
the level of its particularity. Hence, all the types of created
species not indicated as observables, become intractable. For
instance, the observable \texttt{\%obs: Fus3PPFus3(T180\textasciitilde p,Y182\textasciitilde p)}, which is a double phosphorylated MAPK kinase Fus3, is associated with 14 rules of the following form:

\begin{itemize}
\item \texttt{Fus3(dock!1,T180\textasciitilde p,Y182\textasciitilde
    p),Sst2(S539,mapk!1) \\
    -> Fus3(dock,T180\textasciitilde p,Y182\textasciitilde
    p),Sst2(S539,mapk)}

\item \texttt{Ste7(ste5!2,S359\_T363\textasciitilde
    pp,mapk!1),Ste5(ste7!2), Fus3(dock!1,T180\textasciitilde
    p,Y182\textasciitilde p) \\
    -> Ste7(ste5,S359\_T363\textasciitilde
    pp,mapk),Ste5(ste7), \\
    Fus3(dock,T180\textasciitilde p,Y182\textasciitilde p)}

\item \texttt{Fus3(dock!1,T180\textasciitilde p,Y182\textasciitilde
    p),Ste11(mapk!1) \\
    -> Fus3(dock,T180\textasciitilde p,Y182\textasciitilde p),Ste11(mapk)}
\item
  \texttt{Ste5(ste7!1),Ste7(mapk!2,ste5!1,S359\_T363\textasciitilde
    pp), Fus3(dock!2,T180\textasciitilde p,Y182\textasciitilde u) \\
    -> Ste5(ste7!1),Ste7(mapk!2,ste5!1,S359\_T363\textasciitilde pp), \\
    Fus3(dock!2,T180\textasciitilde p,Y182\textasciitilde p)}
\item \texttt{...etc.}
\end{itemize}

However, as it is with the model specification, as it is infeasible to observe all potentially important variations of species, we have to resort to what we know we want to observe.

Therefore, the considered dataset consists of standard 31 variables, patterned after the original paper. There is also an extended 977 variable set but it has yet to be explored with parallel computations. This number was dictated by the snapshot of all existing species at the pick of the simulation (\textasciitilde 1,000 sec after the stimulus appearance) used then as a list of observables in the simulation.

\subsubsection{Perturbed model}

To compare the outcome of applied methods, the model was simulated in two states, which are called here ``perturbed'' and ``unperturbed''.
By the unperturbed model we call the standard ``wild type'' pathway dynamics. The perturbed one relates to an experimentally observed phenomenon of combinatorial inhibition. It occurs when the copy-number of protein scaffold is largely increased and impossible to fully assemble the complex that doubly activates Fus3 because all available members of the complex are used up on too many scaffold proteins.

\subsubsection{Simulator}
\label{sec:simulator}
Models written in Kappa language are supported by KaSim simulator. By default, reaction rates are computed applying the law of mass action~\cite{Chylek2014} but can easily be adjusted to follow any kinetic law (e.g.\ Michaelis-Menten, Hill's Law). What can be found under the hood is a direct particle-based variant of Gillespie's method. A general version of Gillespie’s method, also called exact stochastic simulation algorithm (SSA) or kinetic Monte Carlo is a common simulation method for modelling time-evolution of stochastic chemical reaction systems. Numerical stochastic simulations are known to be computationally intensive and a lot of efforts have been made to improve their efficiency~\cite{Gillespie2007}. The most popular and effective solution, implemented in KaSim, is called “network-free” because rules transforming reactant into products are applied directly at runtime to advance the state of a system. As a result, it does not have to generate the full reaction network beforehand and is therefore independent of its size~\cite{Hogg2014}.

		\begin{figure}
		\includegraphics[width=\textwidth]{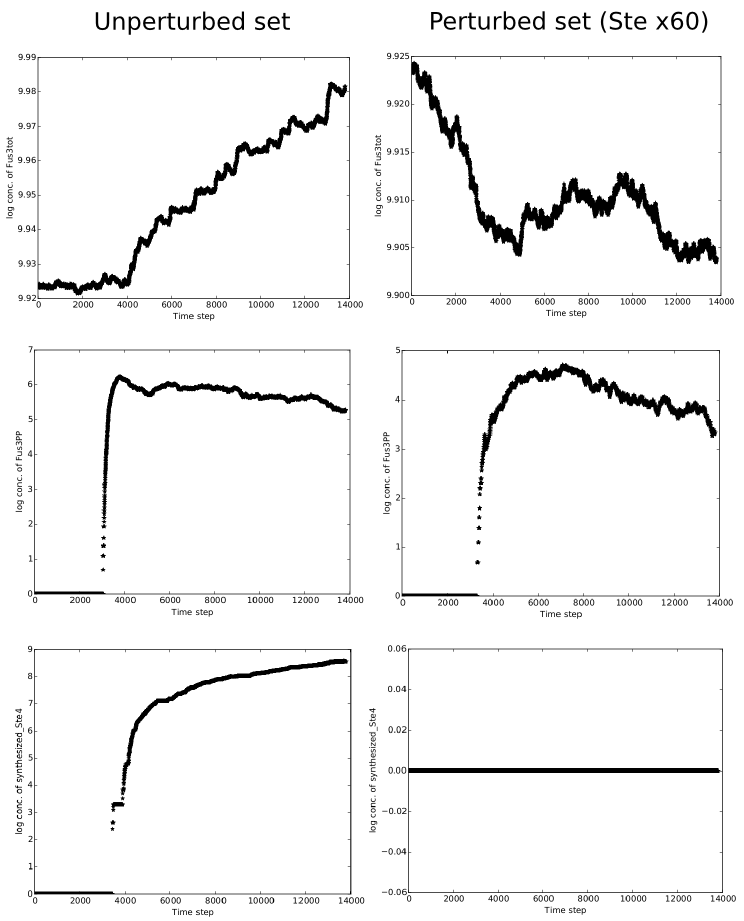}
		\caption{Model dynamics in unperturbed and perturbed states for characteristic protein species. The perturbed ensemble model showed a decrease in Fus3 activation (Fus3PP) being a key observation of the combinatorial inhibition. As we can see, the synthesis of St4, which happens in the nucleus, was inhibited under perturbed state (plot with a flat line).}
		\label{fig:dynamics}
		\end{figure}

\section{Applied methods and results}
	\subsection{Correlation Explanation}

		\subsubsection{Choice rationales}
		
\label{sec:corexmotivation}

Since a rule representation may vary in generalization, it can be applied to more then one reaction that satisfies it. In other words rules serve as the reaction and species generator. It results in the unpredictability of species types and their importance emerging over time. Especially, if the model is of a large magnitude.

On the other hand, the intrinsic modularity of Kappa syntax opens the path to large integrative models, gradually assembled from the collections of reusable rule-based syntactic modules.

However, models are currently built in a fully controllable and stringent fashion. It leaves the notion of modularity and its experimental aspect risky and unexplored. Thinking ahead, the rule notation can be understood as an updatable, machine readable and executable knowledge representation and storage, replacing the usually manual revision of papers required in the model construction~\cite{Agnes2014}. We could allow for uncontrollable, collective model growth in a form of rule stacks and then verify inner links and hierarchies in the system. That could guide an automatic trimming of the model size. Hence, the question is whether and how we could restrict a model to only these rules which are most informative.

Likewise the question lies what exactly does it mean to say a species is ``important'' or ``informative'' and what do groups of biologically important species share with each other? Are they strongly interlinked modules of the system? Could they guide the rule-based modularity idea? The last question is especially important, since the scope of elementary parts of RB model is not yet clear. Are these three, four, five reaction rules? Is there any other measure of mechanisms granularity?

Facing these kind of questions, we opted to test one of recently realised methods Correlation Explanations (CorEx) that applies information theoretic objective to learn a hierarchy of more abstract representations of the data.

Having the hierarchy of latent variables, we can pose more precise questions, such as:

\begin{itemize}
\item
What subset of species has common underlying dynamics?
\item
How strong is the correlation between species grouped under the same latent variable?

\item
How many underlying hidden causes can be identified out of the observed high-dimensional species dynamics?
\end{itemize}

and finally:
\begin{itemize}
\item
What could be the meaning of these ``hidden causes'' for molecular signalling?
\end{itemize}

The algorithm was previously applied in a biological context for identification of targets for a cancer therapy~\cite{Pepke2015}. Furthermore, a similar method mentioned in CorEx paper, called the information bottleneck, was previously applied for trimming of gene ontology (GO)~\cite{Jin2010} to compress the data into a smaller representation. In contrast, in CorEx the redundant information is preserved ignoring uncorrelated random variables~\cite{NIPS2014_5580}.

		\subsubsection{Method description}
		
\newcommand{\tc}{TC}
\newcommand{\argmax}{\arg\!\max}

In this section, we discuss an information theoretic approach for building a model on dynamics of the species concentrations. This method, proposed recently for a general domain~\cite{NIPS2014_5580,steeg2015corex_theory}, learns a hierarchy of latent variables that maximally inform correlation between the observed species dynamics. Herein, correlation refers to mutual information between a set of variables, and not just a linear correlation.

Before we delve into the details of the method for our specific settings, it should be noted that we disregard the time series nature of species copy-number dynamics in this method application.

Let $G$ be a set of random variables representing copy-numbers for all the species. Then, $X_G$ is a joint random variable on $G$. For a species $i$, all the copy-number values of the time series are assumed to be independent samples of a random variable $X_i$. As such, we can see that there is a contradiction since consecutive samples in the time series would have a correlation~(not i.i.d.). For obtaining uncorrelated samples, one can take sub-samples of the time series, either at uniform interval or using any other relevant technique.

Following the notations in \cite{NIPS2014_5580}, total correlation $\tc(.)$ between a set of variables $X_G$ is defined as below.

\begin{align}
&
\tc(X_G)
=
\sum_{i \in G}
H(X_i)
-
H(X_G)
\\
&
\tc(X_G)
=
I(X_1;\cdots;X_g)
\end{align}

Here $H(X_i)$ is entropy on a random variable $X_i$; and $H(X_G)$ is a joint entropy on $X_G$. Another interpretation of $\tc(X_G)$ is that it is mutual information, $I(.)$, between all the variables in the set $G$. Typically mutual information is expressed between a pair where each element of the pair can be a set of random variables. Here, we are instead expressing mutual information between a $g$ dimensional triplet of random variables, where $g$ is a number of random variables in the set $G$.

In our problem of learning a model of species dynamics, evaluating mutual information~(or total correlation) between all random variables would not be of much value. We are instead interested in evaluating mutual information between some subsets of species. However there are two problems along these lines: i) we do not know for which subsets of species we should evaluate mutual information and there can be a large number of permutations to explore~(depending on the size of a subset and the $G$ set); ii) non-parametric estimation of mutual information between random variables is a hard problem~\cite{kraskov2004estimating,suzuki2008approximating,pal2010estimation,shuyang2015MI}.
To tackle these, we formulate our algorithm such that; i) we assume the individual species copy-number variables $X_i$ to be Gaussian; however, we do not assume that \emph{the set of variables} has to be Gaussian~(the later is a stronger assumption); ii) we are interested in only those subsets where variables have low mutual information conditioning on a latent variable~(or high mutual information between variables explained by a latent variable).

Along these lines, let us define a new information theoretic quantity $\tc(X_G;Y_F)$.
\begin{align}
\tc(X_G;Y_F)
=
\tc(X_G)
-
\tc(X_G|Y_F)
\end{align}
$\tc(X_G;Y_F)$ is a total correlation~(or mutual information) in the set of random variables $X_G$ explained by a set of latent variables $Y_F$. $\tc(X_G|Y_F)$ is a total correlation between the random variables $X_G$ that can not be explained by $Y_F$, i.e. conditional total correlation~(conditional mutual information). If the latent variables $Y_F$ can explain the total correlation in $X_G$ perfectly, then $\tc(X_G|Y_F)=0$. Ideally, we would like to learn $Y_F$ if exists. Thus intuitively, optimal $Y$ would correspond to minimum of $\tc(X_G|Y_F)$. In our formulation, we can express optimization of $Y_F$ as optimizing conditional distributions $P_{Y|X}$.

Let us first consider optimization of a single latent variable $Y$, and then generalize it later.

\begin{align}
\argmax_{Y:p(y|x_G)}
\tc(X_G;Y)\ s.t.\ |Y|=k
\end{align}
Here $Y$ is a discrete random variable; $x_G$ is a sample of the random variable $X_G$ and $y$ is sample of $Y$. We optimize $Y$ by learning the conditional distribution $p(y|x_G)$. Now, we further extend it for multiple latent variables, where each latent variable explains total correlation in a subset of the species concentration variables.

\begin{align}
\argmax_{G_j, p(y_j|x_{G_j})}
\sum_{j=1}^m
\tc(X_{G_j}; Y_{j})
\end{align}

We have introduced $m$ latent variables here with $Y_j$ explaining correlation between random variables in the corresponding subset $G_j \subset G$. Here these subsets can have an overlap.

Optimizing the above objective function seems hard. However, as explained in detail in \cite{NIPS2014_5580,steeg2015corex_theory}, it can be solved very efficiently for practical purposes. We omit these optimization details and refer readers to the original papers introducing this algorithm for the first time~\cite{NIPS2014_5580,steeg2015corex_theory}. Computational complexity of the method is linear with respect to the number of samples and number of variables in the set $G$. Furthermore, as an unsupervised method, it requires no assumption about the learned model. The code implementation for this algorithm is publicly available from the original authors~\footnote{https://github.com/gregversteeg/CorEx}.


		\subsubsection{Results}

The CorEx algorithm was applied both to perturbed and unperturbed datasets and yielded two results with a single layer of hidden variables. In both cases, presented results were the maximal values the data sets could be divided to. Further increase of the number of hidden units did not change their values.

\begin{sidewaysfigure}
		\includegraphics[width=\textwidth]{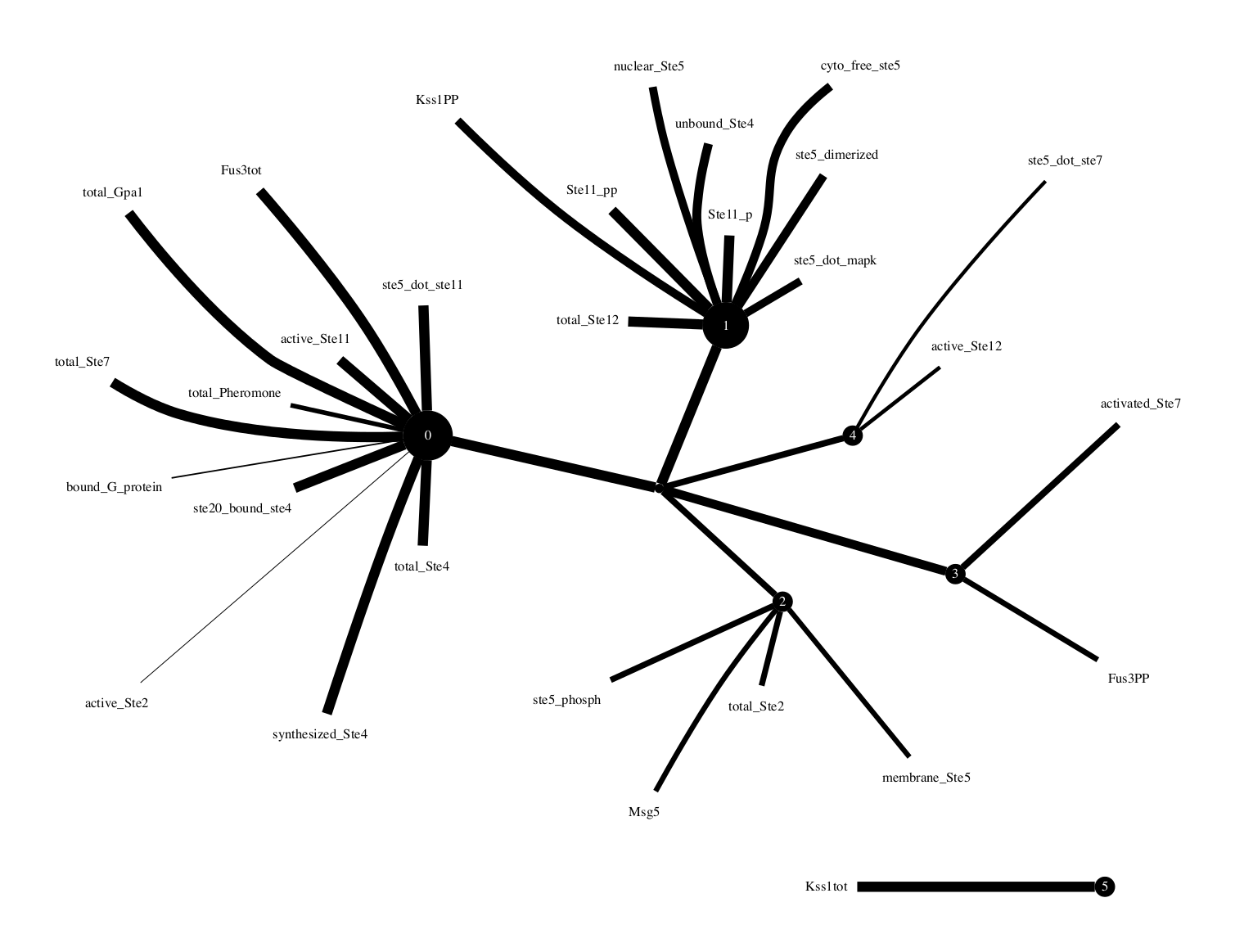}
		\caption{Result of 31-variable dataset \textbf{without} perturbation. Intrinsic dimensionality was found to be 6. Variable numbers are shown in the middle node of each group. Edge weights leading from a group centre to its member are dictated by its explanatory contribution to remaining group members}
		\label{fig:unpert}
		\end{sidewaysfigure}

\begin{sidewaysfigure}
		\includegraphics[width=\textwidth]{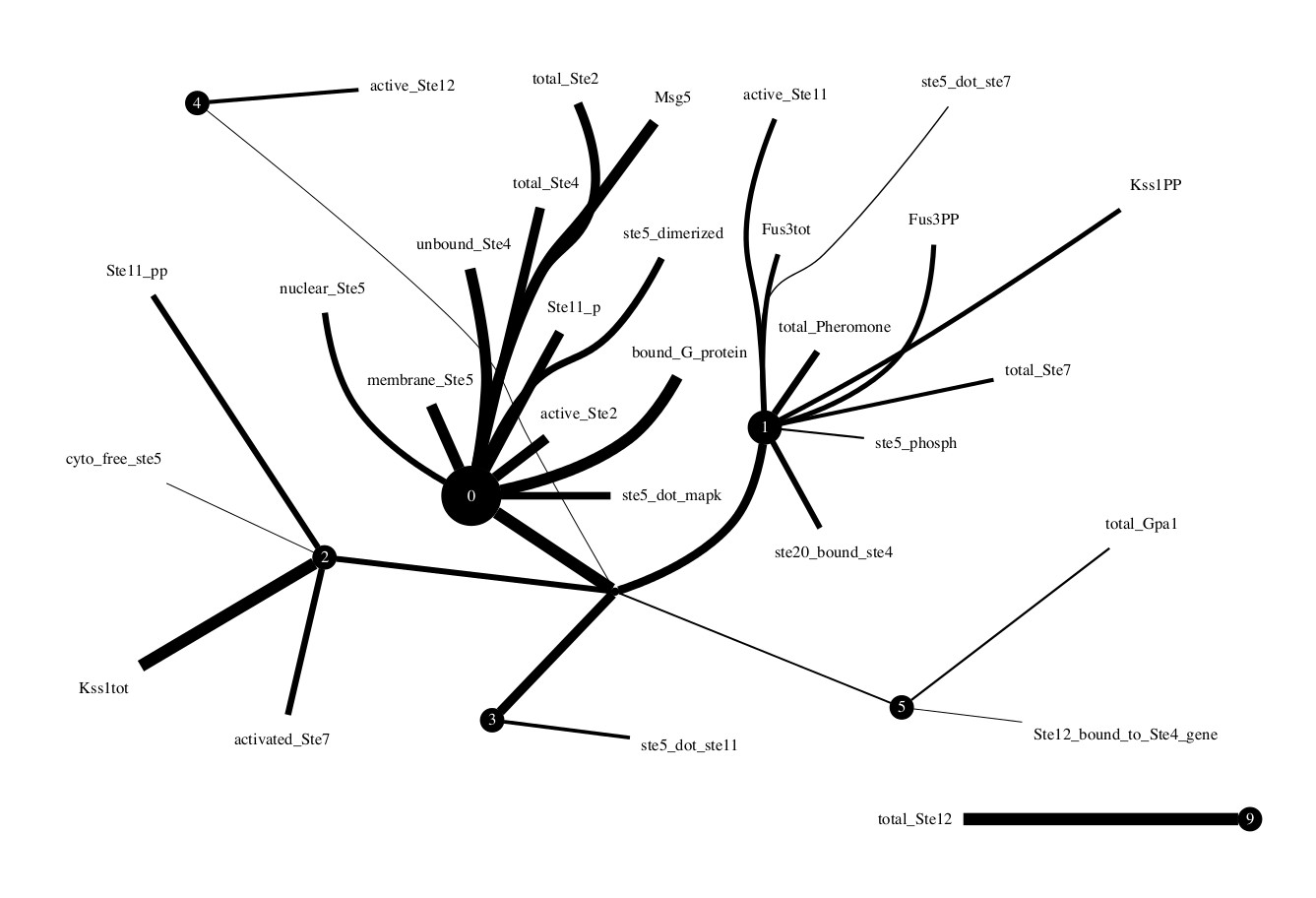}
		\caption{Result of 31-variable dataset \textbf{with} perturbation. Intrinsic dimensionality was found to be 6. Variable numbers are shown in the middle node of each group. Edge weights leading from a group centre to its member are dictated by its explanatory contribution to remaining group members}
		\label{fig:pert}
              \end{sidewaysfigure}

For both sets [Figures~\ref{fig:pert} \&~\ref{fig:unpert}] CorEx found 6 latent variables, where 8-9 out of 31 biologically plausible variables were expected.Biologically plausible variables were thought to be all these observables that contained Ste5 scaffold protein, known to be a nucleation point of the system~\cite{Suderman2013}. However, the preliminary intuitions did not align perfectly with the algorithm results.

In the unperturbed data tree we can distinguish two important groups, `0' and `1'. They are recognisable in the perturbed data tree as they preserved half of their members from the former set. However, contrary to the unperturbed data tree, where the members of both groups have similarly balanced strong relations, the perturbed set shows far uncertain correlations, mostly concentrated in the group `0'.

		\subsubsection{Interpretation and analysis}

The interpretation of the results was conducted on two levels. The first one is based on the biological knowledge about the process. The second one is supported by the dynamic analytical tools provided by the KaSim simulator.

\begin{figure}
		\includegraphics[width=\textwidth]{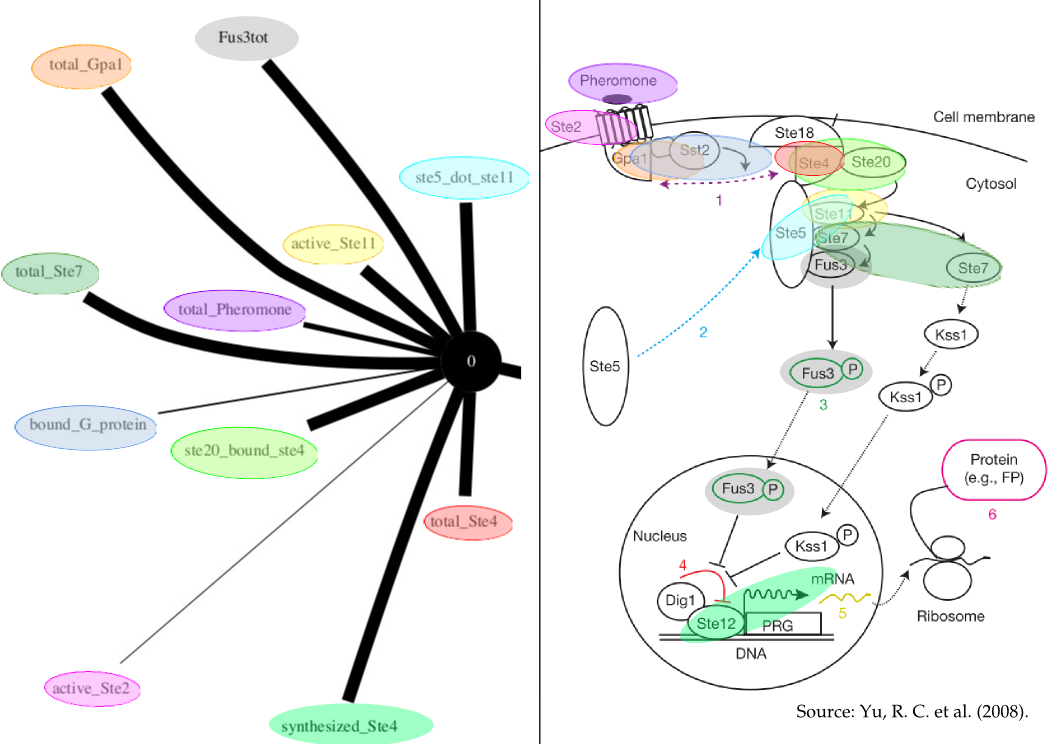}
		\caption{On the left, a fragment of unperturbed data-tree with Group `0'. On the right, the process diagram for a comparison. A biological interpretation of the strongest group `0' refers to the most important steps indicating critical events in the successful signal propagation. As the authors argued, the assembly of decamer involving Ste5 dimerization does not belong to the most crucial events guaranteeing the signal transfer.}
		\label{fig:biounpert1}
		\end{figure}

\begin{figure}
		\includegraphics[width=\textwidth]{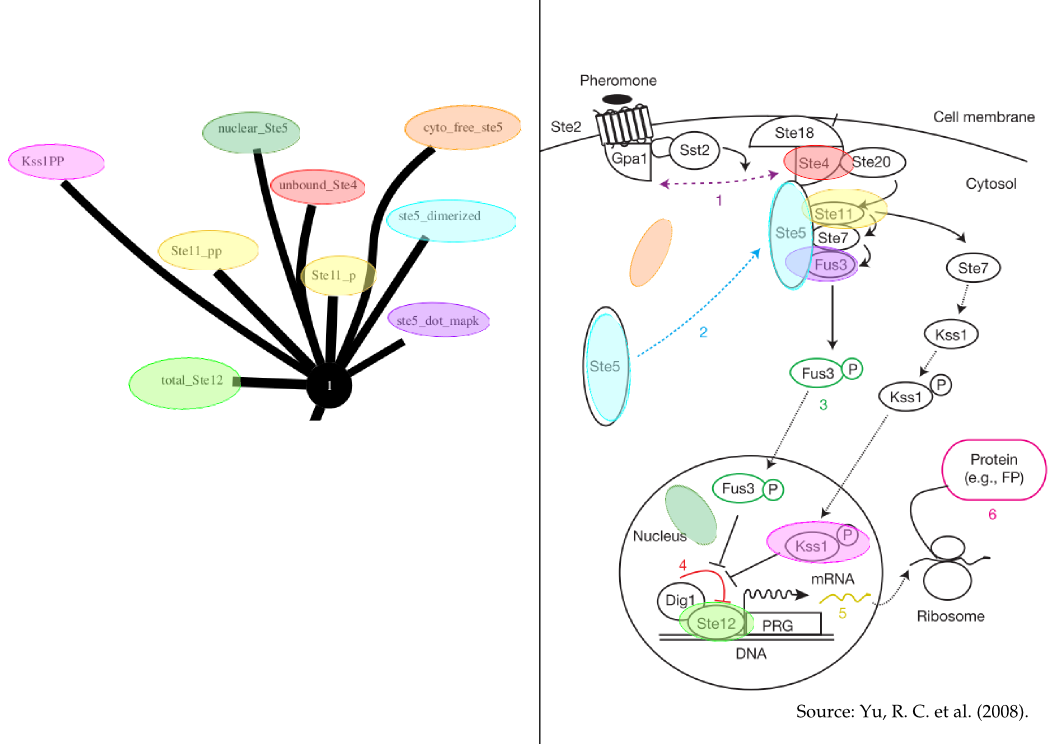}
		\caption{On the left, a fragment of unperturbed data-tree with Group `1'. On the right, the process diagram for a comparison. The second highly scored group indicates less vital events, related to dimerization, and the impact of Kss1 kinase on the Ste4 activation.}
		\label{fig:biounpert2}
		\end{figure}

\begin{figure}
		\includegraphics[width=\textwidth]{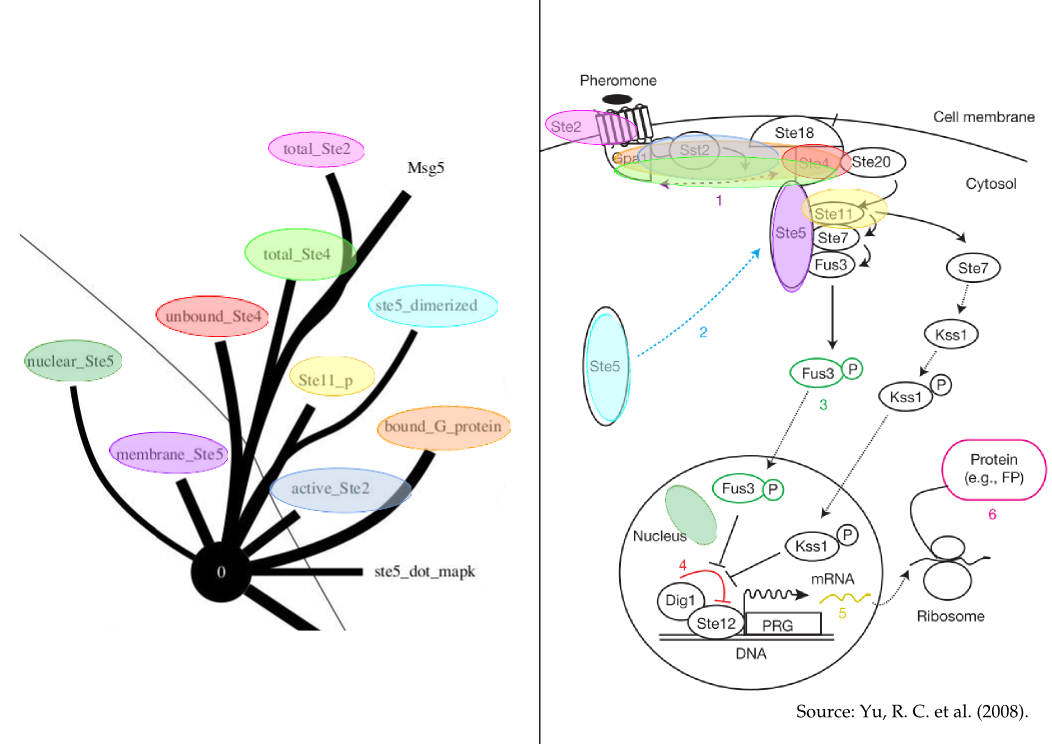}
		\caption{On the left, a fragment of perturbed data-tree with Group `0'. On the right, the process diagram for a comparison. The strongest group indicates the earliest events located upstream to the Fus3 poshorylation (observable called Fus3PP), preceding the complexation step}
		\label{fig:biopert1}
		\end{figure}

\begin{figure}
		\includegraphics[width=\textwidth]{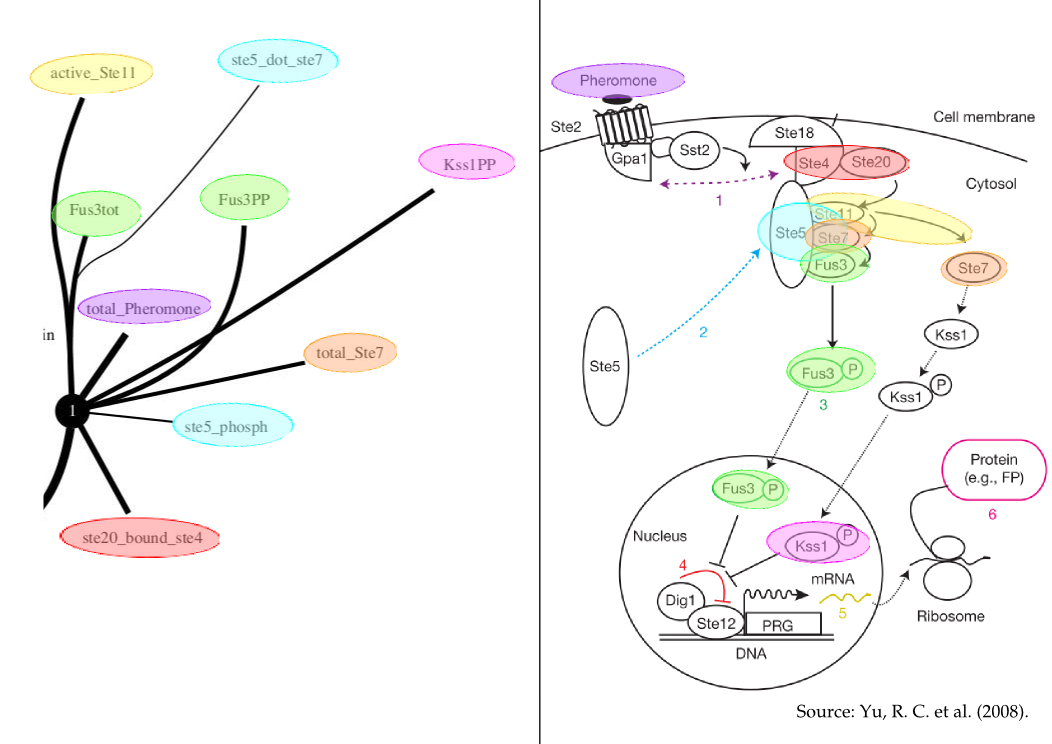}
		\caption{On the left, a fragment of perturbed data-tree with Group `1'. On the right, the process diagram for a comparison. The second strongest group of perturbed data tree reflects weak correlation between members and the unsuccessful activation of transcription factor St4.}
		\label{fig:biopert2}
		\end{figure}

Generally speaking, the CorEx algorithm successively subsets data into a defined number of latent variables guided by species dynamics. Results appears to be consistent with the differences between perturbed  [Figures~\ref{fig:biopert1}~\&~\ref{fig:biopert2}] and unperturbed models [Figures~\ref{fig:biounpert1} \&~\ref{fig:biounpert2}]. The group ordering, referring to the strength of inter-correlations, shows which event takes the lead in two cases.
Earliest events upstream to the formation of signalling cascade appeared to be the leading ones in the perturbed simulation.
This is consistent with the fact that phosphorylation of Fus3 kinase distinctively drops when the amount of Ste5 protein scaffolds competing for binding kinases increases [Figure~\ref{fig:dynamics} in Section~\ref{sec:simulator}].
As the Fus3 phosphorylation was not entirely blocked, the second latent variable relates to events leading to Fus3 phosphorylation. Thus it is more consistent with the group `0' apart from transcription in the nucleus, which was inhibited in the perturbed data.


Owing to the static causal analysis provided by the simulation software, we can ask whether important observables relate to frequently executed rules. The most powerful visualization output is a flux map, which tracks the overall influence of rule applications on each other~\cite{Feret2012}. It is a directed and weighted graph with rules as vertices and edges annotated with positive or negative weights [Figure~\ref{fig:fluxmap}]. Dependent on simulation parameters (selected time or number of events), a flux map might vary in structure (for details of our simulation parameters see section~\ref{sec:datasets}).

\begin{sidewaysfigure}
		\centering
                \includegraphics[width=\textwidth]{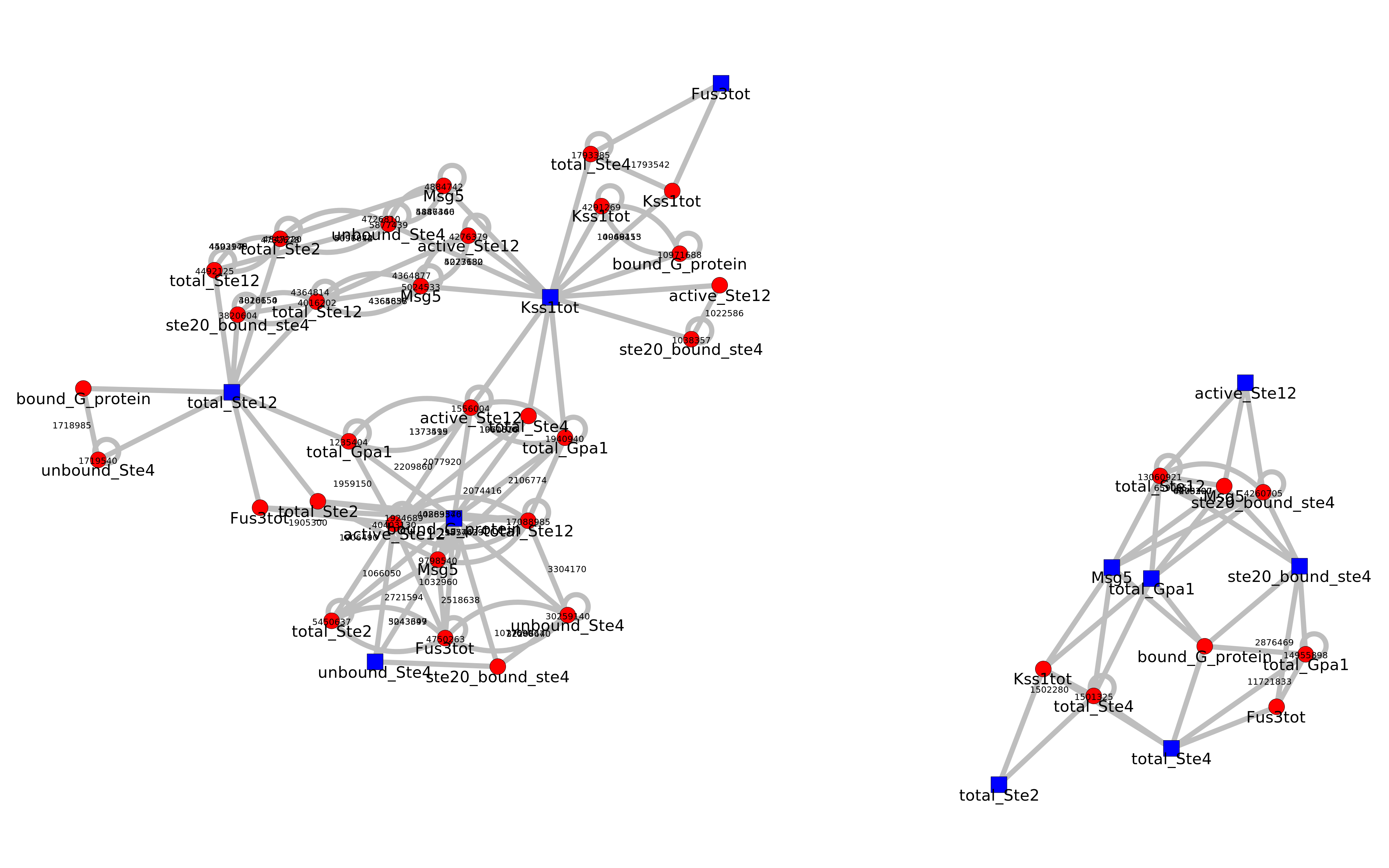}
		  \caption{An example of two flux map subgraphs mapping observables to related rules. The perturbed dataset with weights > 1,000,000, blue nodes denote observables, red nodes rule names.}
		  \label{fig:obs2rule1}
               \end{sidewaysfigure}

\begin{sidewaysfigure}
		  \flushleft
		  \includegraphics[width=\textwidth]{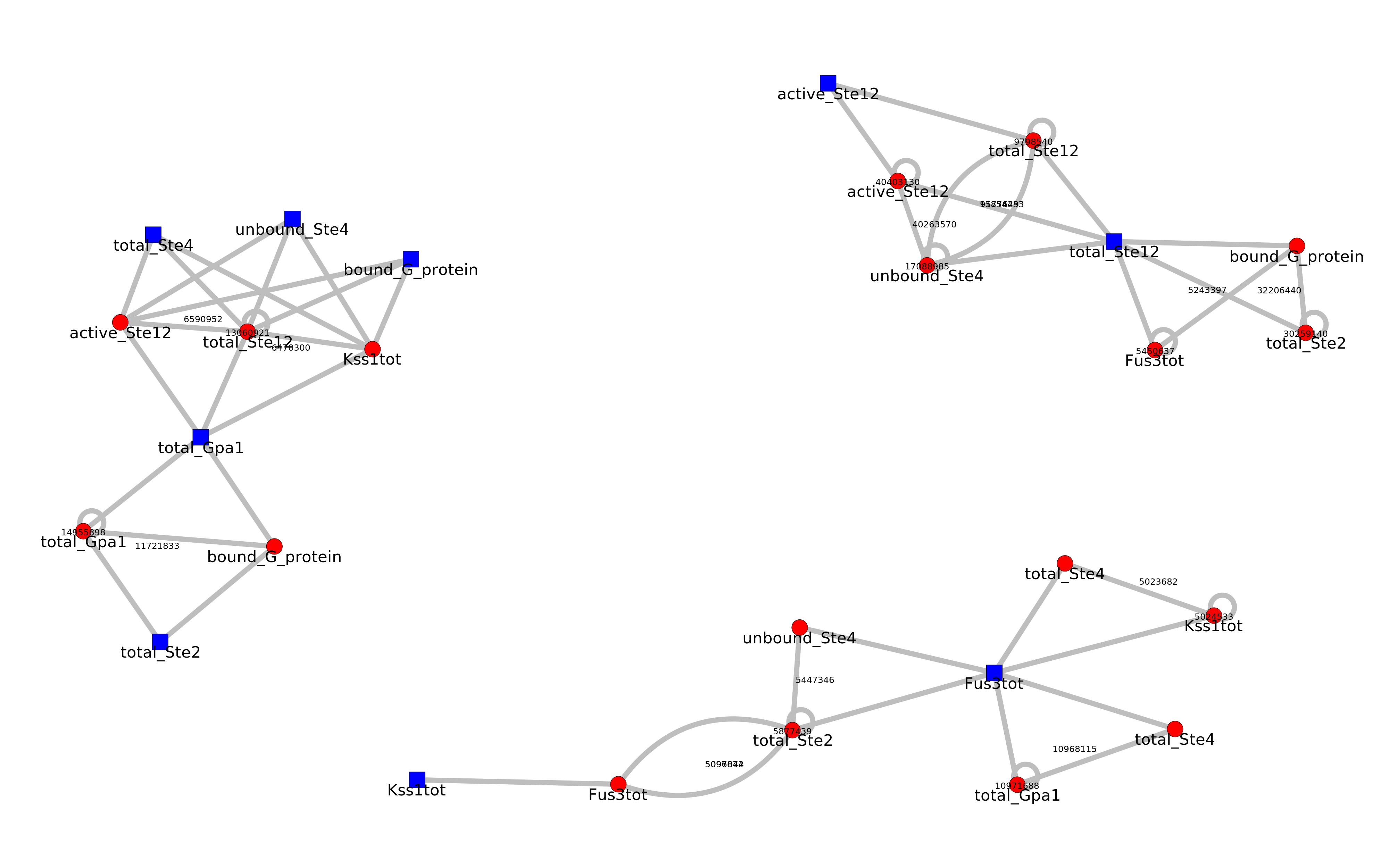}
		  \caption{An example of two flux map subgraphs mapping observables to related rules. The perturbed dataset with weights > 5,000,000, blue nodes denote observables, red nodes rule names.}
		  \label{fig:obs2rule2}
		\end{sidewaysfigure}

Both untrimmed graphs for unperturbed and perturbed models had 233 vertices but they differ from each other in the edge number (unperturbed- 2,753, perturbed- 2,422). Weights range from 0 to 407,172,203. An important note is that vertices are rules. Hence, to compare them with the output of CorEx, thus subsets of observables, first observables had to be mapped to rules they referred to [Figure~\ref{fig:obs2rule1} \&~\ref{fig:obs2rule2}]. The weight cutoff varies with inverse proportion to the number of observables in flux map subgraphs. Therefore, we compared  different subgraphs by gradually removing less and less vertices given a set of thresholds for weight values.  The aggregated results are presented in the Figure~\ref{fig:fluxmap}. As we can observe, the frequency of rule application relates to subsets obtained with the CorEx algorithm but cannot explain them fully.

We stated some questions in the Section~\ref{sec:corexmotivation}, which we would like to comment on or even answer to in the following part.
We have learned that the algorithm used on time series datasets divided the species into most important ones over the entire course of time series, with results depended from the outcome of signalling process.
Hence, it did not inform us about intrinsic modularity of the system, what would relate to more ``horizontal'' division of time courses (when looking at the process diagram). Perhaps the considered system is far too small thus interlinked to observe invariable modules among species (encapsulations). Hence, the result might be then more correctly named as a form of ``compression''. Furthermore, given the limited number of experiments and the model size and its character, we are not yet ready to precisely answer the question of biological meaning related to the importance and informativeness indicated by the algorithm.

		\begin{figure}
		\includegraphics[width=\textwidth]{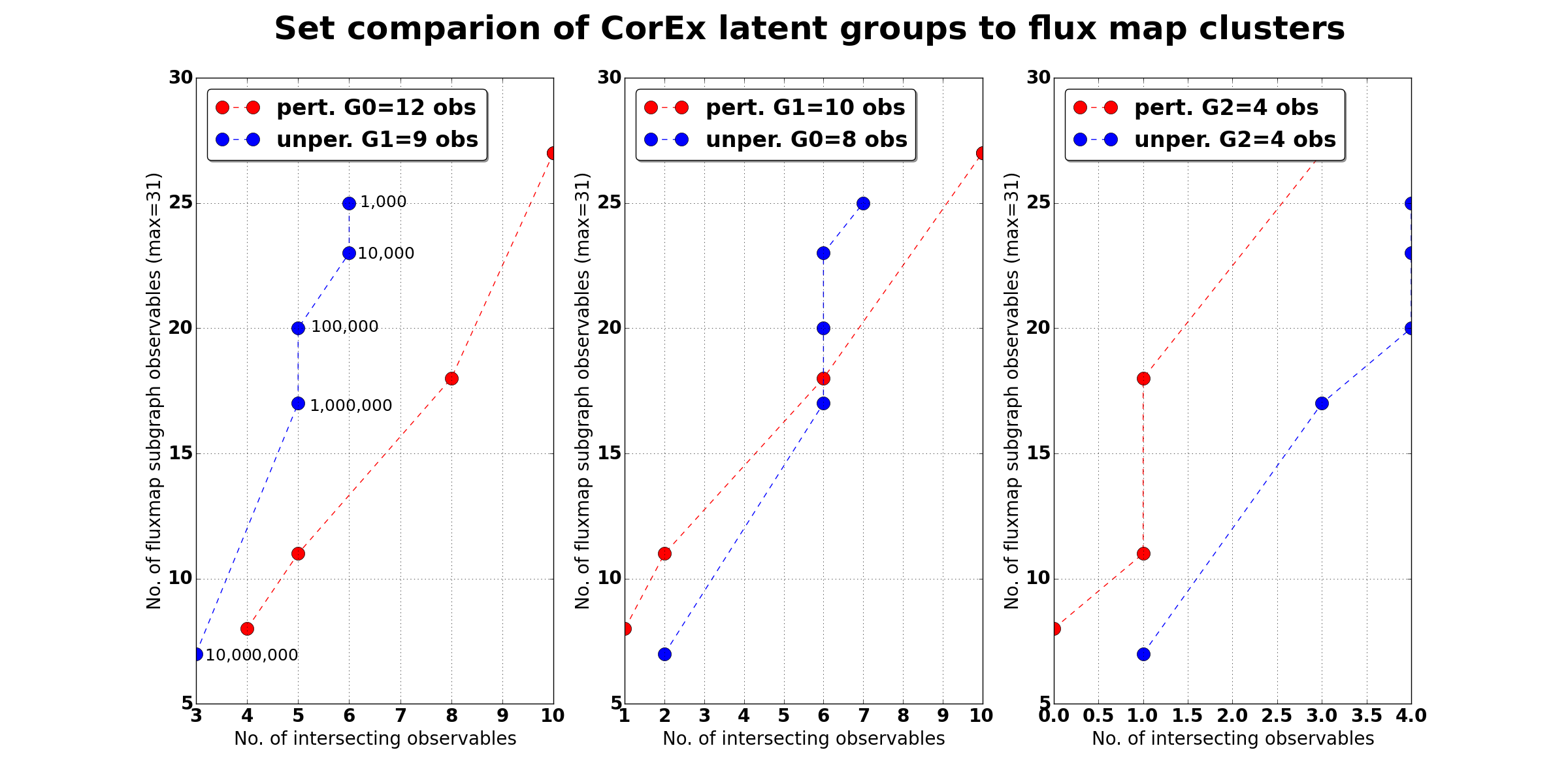}
		\caption{Three top-scored groups of latent variables (G0, G1, G2) found with CorEx, both for perturbed (red) and unperturbed (blue) simulations and five flux map subgraphs with weights above (starting from left on x-axis) 10 000 000, 1 000 000, 100 000, 10 000, 1 000 units (two last sets in the perturbed set overlap). The comparison of changes in the number of intersecting observables with decrease of stringency in rule importance shows that perturbed system gives seemingly higher overlap between compared groups than the non-perturbed dataset.}
		\label{fig:fluxmap}
		\end{figure}

		\begin{sidewaysfigure}
		\includegraphics[width=\textwidth]{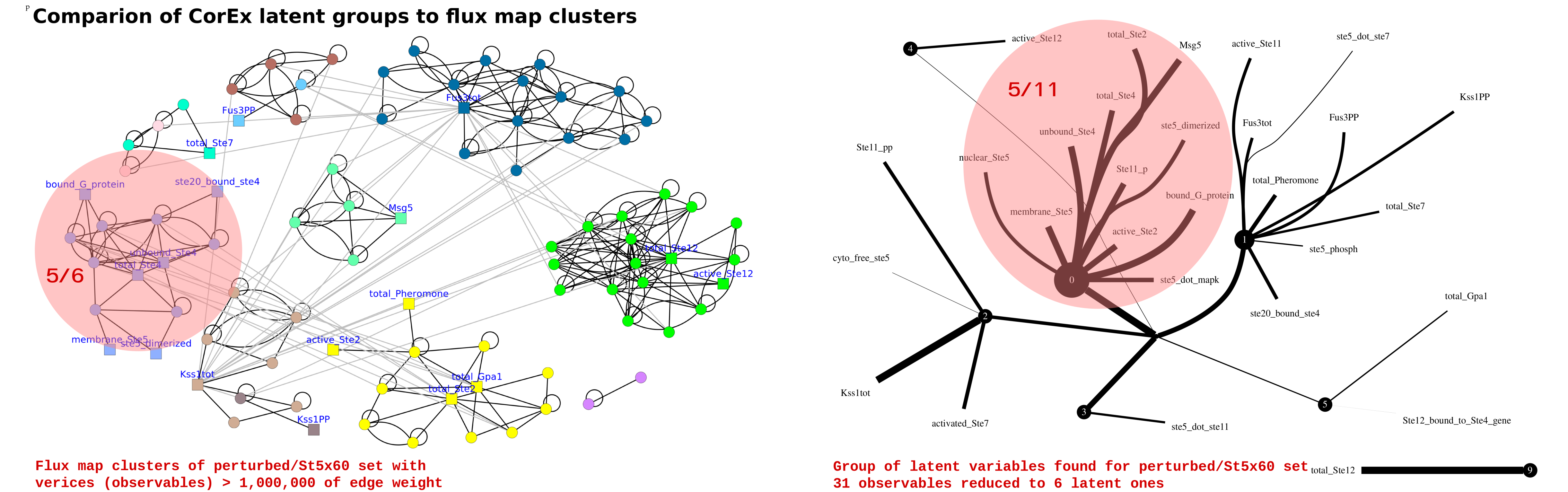}
		\caption{The largest intersection between the network of most frequently executed rules and the latent groups is apparently more visible in the perturbed model, which confirms a lack of coherence in species behaviour.}
		\label{fig:fluxmapclust}
              \end{sidewaysfigure}

	\subsection{Chaos Time Series Analysis}

To compare results with the outcome of CorEx algorithm and discover other aspects hidden in our data, we applied methods of nonlinear time series analysis \cite{hegger1999practical}. Similarly, we used both the perturbed and unperturbed datasets (for more details about used datasets see Section \ref{sec:datasets}). 
To bypass an obvious division into a pre- and post-pheromonal stimulation, we cut the beginning 1,000 sec and used only the part after the stimulation. Furthermore, to cap the computation time, we cut the data from original 10,801 (three events per second) observations to 3,600 (one event per second).

First we examined our data by creating recurrence plots for dynamical systems \cite{eckmann1987recurrence}. The recurrence plot is an array of dots in a $N x N$ square, where a dot is placed at $(i,j)$ whenever $x(j)$ is sufficiently close to $x(i)$. For the purposes of this study we selected an embedding dimension of 10 and time delay 5 to keep the computational time within reasonable limits.

In general, the recurrent plot shows the times at which a phase space trajectory visits roughly the same area in the phase space \cite{Marwan2008}. The authors \cite{eckmann1987recurrence} defined small and large scale patters, textures and topologies respectively, to ease their interpretation.

\begin{figure}
\begin{minipage}[b]{0.45\linewidth}
\centering
\includegraphics[width=\textwidth]{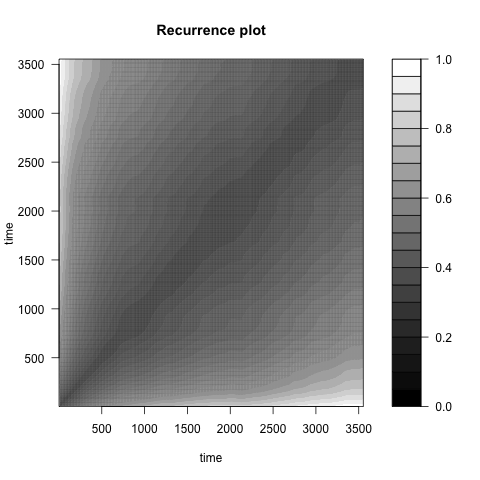}
\caption{Unperturbed model}
\label{recurrence_plot02}
\end{minipage}
\begin{minipage}[b]{0.45\linewidth}
\centering
\includegraphics[width=\textwidth]{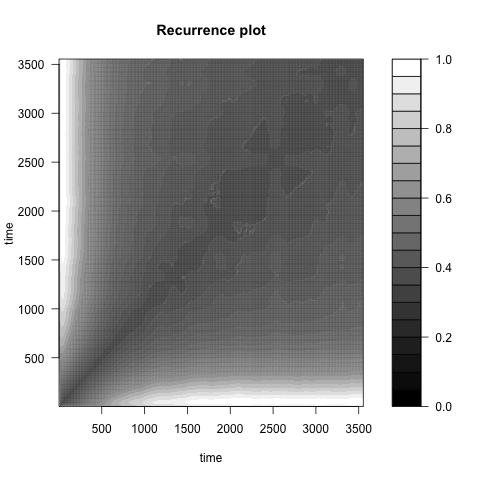}
\caption{Perturbed model}
\label{recurrence_plot01}
\end{minipage}
\end{figure}

The resulting figures [\ref{recurrence_plot01} \& \ref{recurrence_plot02}] are densely grey without distinctive textures or patterns. However, the unperturbed set is seemingly brighter away from the diagonal and distinctively darker along it. This gradient is interpreted as the occurrence of a progressive decorrelation at large time intervals involving a linear trend or drift. The perturbed model presents dynamics pushed a bit more towards randomness.

Next, we created plots, showing the average mutual information index (AMI) of a given time series for a specified number of lags \cite{fraser1986independent}.

\begin{figure}
\begin{minipage}[b]{0.45\linewidth}
\centering
\includegraphics[width=\textwidth]{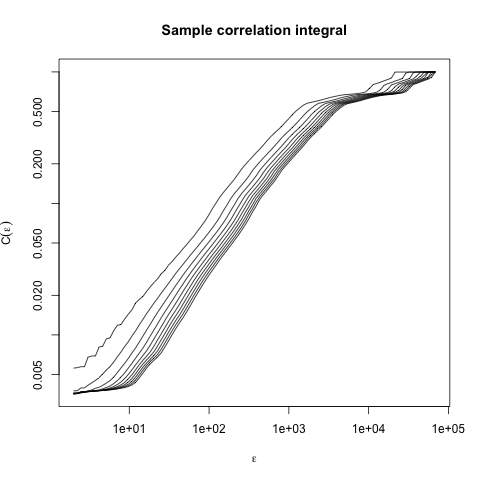}
\caption{Unperturbed data}
\label{Sample correlation integral plot 1}
\end{minipage}
\begin{minipage}[b]{0.45\linewidth}
\centering
\includegraphics[width=\textwidth]{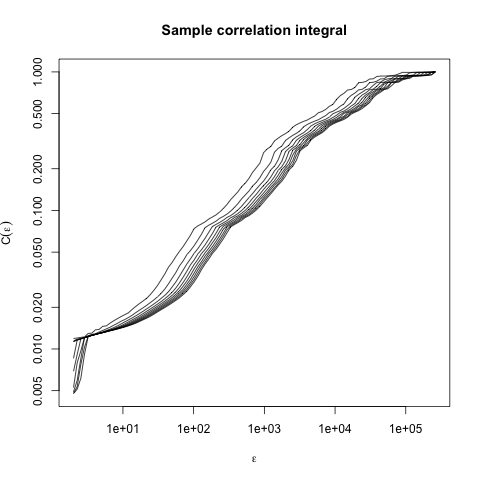}
\caption{Perturbed data}
\label{Sample correlation integral plot 2}
\end{minipage}
\end{figure}

The larger time lag the smaller is the value of AMI. In case of Figure \ref{Sample correlation integral plot 1} AMI drops down almost diagonally, as opposed to the Figure  \ref{Sample correlation integral plot 2}. Thus, the perturbed model is far more unpredictable, showing randomized dynamics and less interdependent relation between events.

\begin{equation}
S=-\sum_{ij}p_{ij}(\tau)ln\frac{p_{ij}(\tau)}{p_{ij}}
\end{equation}

\begin{figure}
\begin{minipage}[b]{0.45\linewidth}
\centering
\includegraphics[width=\textwidth]{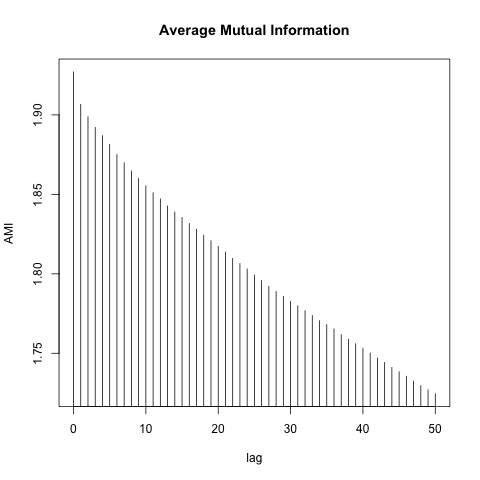}
\caption{Unperturbed data}
\label{}
\end{minipage}
\begin{minipage}[b]{0.45\linewidth}
\centering
\includegraphics[width=\textwidth]{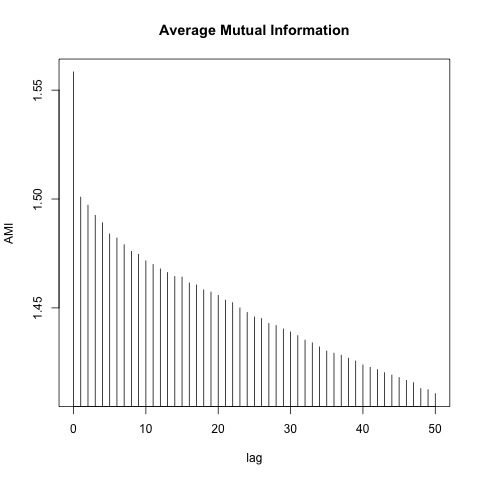}
\caption{Perturbed model}
\label{average_mutual_information_plots}
\end{minipage}
\end{figure}

Next, we created a sample correlation integral plot  \cite{hegger1999practical}. The correlation integral can be approximated by the correlation sum. The correlation sum counts the number of pairs $(\overrightarrow{x}(i),\overrightarrow{x}(j))$ in a given set of vectors that are at most $\epsilon$ apart.

\begin{equation}
C(\epsilon)=\frac{1}{N(N-1)} \sum_{i=1}^N \sum_{j=i+1}^N \Theta (\epsilon-\Vert \overrightarrow{x}(i) - \overrightarrow{x}(j)\Vert), \overrightarrow{x}(i) \in \mathbb{R}^m
\end{equation}

As the perturbation involved a single parameter and demonstrated naturally occurring phenomenon (not randomised), differences between these two plots were not expected to be extreme. Nonetheless, the results are coherent both with the understanding of process and the CorEx algorithm. However, for our purposes, these methods present a more distanced view on the system dynamics, missing a decoupling problem of individual species relations.

\section{Conclusions}

Overall, this project offered a fruitful chance for an exploration of multivariate time series analysis. 
We have learned that the approach offered by the CorEx method might be very promising in analysis of rule-based 
models. However, it requires further testing with models that incorporate multiple randomly modified parameters and represent larger advanced processes.
Further, we applied some nonlinear time series methods to our dataset. Though powerful, they offered a bird's-eye view understanding of system dynamics missing species-related details. However, both methods correctly interpreted the process offering a useful insight otherwise inaccessible.

\newpage
\bibliography{references}
\bibliographystyle{plain}

\end{document}